\documentclass[journal]{vgtc}                     

\onlineid{0}

\vgtccategory{Research}

\vgtcpapertype{system}

\title{LLMER: Crafting Interactive Extended Reality Worlds with JSON Data Generated by Large Language Models}

\author{%
  \authororcid{Jiangong Chen}{0000-0002-1890-5110},
  \authororcid{Xiaoyi Wu}{0009-0000-8966-816X},
  Tian Lan, and 
  \authororcid{Bin Li}{0000-0001-6002-677X}
}

\authorfooter{
  \item Jiangong Chen, Xiaoyi Wu, and Bin Li are with the Department of Electrical Engineering at the Pennsylvania State University. Email: \{jiangong, xfw5193, binli\}@psu.edu
  \item
  	Tian Lan is with the Department of Electrical and Computer Engineering at the George Washington University.
  	E-mail: tlan@email.gwu.edu
}

\abstract{
    The integration of Large Language Models (LLMs) like GPT-4 with Extended Reality (XR) technologies offers the potential to build truly immersive XR environments that interact with human users through natural language, e.g., generating and animating 3D scenes from audio inputs. However, the complexity of XR environments makes it difficult to accurately extract relevant contextual data and scene/object parameters from an overwhelming volume of XR artifacts. It leads to not only increased costs with pay-per-use models, but also elevated levels of generation errors. Moreover, existing approaches focusing on coding script generation are often prone to generation errors, resulting in flawed or invalid scripts, application crashes, and ultimately a degraded user experience. To overcome these challenges, we introduce LLMER, a novel framework that creates interactive XR worlds using JSON data generated by LLMs. Unlike prior approaches focusing on coding script generation, LLMER translates natural language inputs into JSON data, significantly reducing the likelihood of application crashes and processing latency. It employs a multi-stage strategy to supply only the essential contextual information adapted to the user's request and features multiple modules designed for various XR tasks. Our preliminary user study reveals the effectiveness of the proposed system, with over 80\% reduction in consumed tokens and around 60\% reduction in task completion time compared to state-of-the-art approaches. The analysis of users' feedback also illuminates a series of directions for further optimization.
} 


\teaser{
  \centering
  \includegraphics[width=\linewidth]{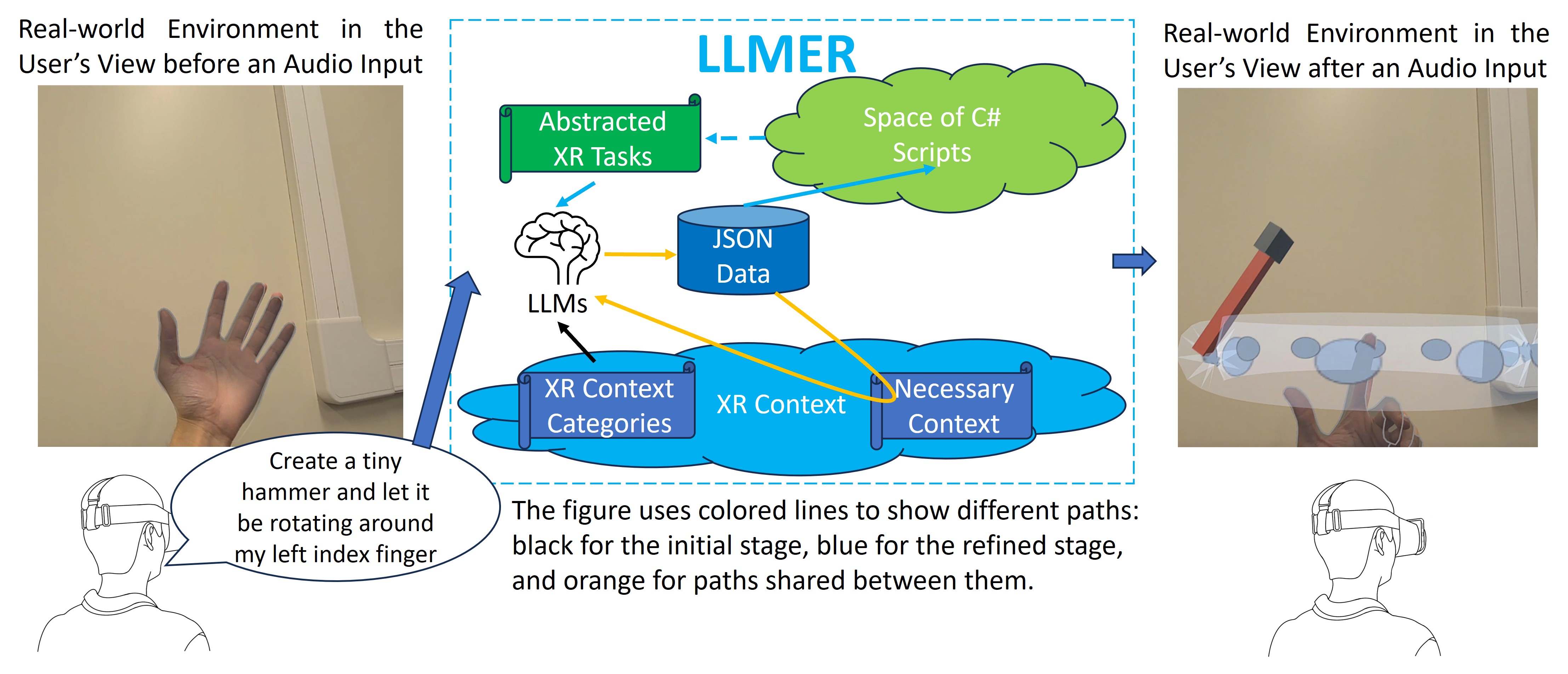}
  \caption{An illustration of our system to enhance the immersive user-XR interaction. It (i) processes language/audio inputs and (ii) leverages LLM to identify the necessary contextual categories for extracting minimal sufficient XR context information and generate interactive conversations with human users as well as structured JSON data -- which can then be executed by various designed modules to create virtual objects and/or animations. This multi-stage approach -- consisting of processing, extraction, generation, and execution, which are empowered by LLM -- effectively projects the complex generation problem into a compact subspace, generates context-relevant structured JSON data in the subspace, and then maps the results back to the original XR space to drive interactions.
  }
  \label{fig:teaser}
}

\graphicspath{{figs/}{figures/}{pictures/}{images/}{./}} 

\usepackage{tabu}                      
\usepackage{booktabs}                  
\usepackage{lipsum}                    
\usepackage{mwe}                       

\usepackage{mathptmx}                  

\usepackage{amsmath}
\usepackage{amssymb}
\usepackage{mdframed}
\usepackage{enumitem}
\usepackage{subfig}

\newcommand{\sysname}{LLMER\xspace}

\begin{document}


\firstsection{Introduction}

\maketitle

The rapid advancement of generative artificial intelligence (GenAI) has revolutionized the creation of various types of content, such as text, scripts, audio, images, and videos. Within this realm, Large Language Models (LLMs) such as GPT-4 have emerged as powerful tools, particularly adept at generating coherent and contextually relevant text and script data. 
On the other hand, Extended Reality (XR) (see \cite{chuah2018and}, an umbrella term encapsulating Augmented Reality (AR), Virtual Reality (VR), Mixed Reality (MR), and everything in between) facilitates interactions among human users, virtual content, and real-world environments. The fusion of LLM with XR technologies (see \cite{chen2024gpt,giunchi2024dreamcodevr,de2023llmr}) seems a natural approach for enhancing such interactions, providing a more effective method than standard hand gestures or pre-defined menus.
For example, users can potentially create virtual objects that adapt to the real-world environment or generate specific animations that interact with their limbs, all by simply articulating their needs through natural language.

Despite these promises, integrating LLM into XR systems to enhance the immersive experience presents several challenges: i) XR environments often encompass thousands of artifacts and parameters -- which could easily overwhelm LLM during generation, in terms of both hindering the accurate capture of useful contextual information and increasing the costs associated with pay-per-use Application Programming Interfaces (APIs); For instance, in a scene with hundreds of objects and each with tens of properties, LLMs may create floating cups not adhering to the surface of a table.
ii) Existing approaches often rely on generating and compiling source codes to achieve desired functionalities, but the codes generated by LLMs are prone to errors, leading to compilation issues and application crashes during runtime;
iii) To mitigate code generation errors, prior solutions apply multiple iterations to produce high-quality scripts, which leads to significant generation latency and reduces system responsiveness.

This leads to several critical research questions: (i) How can we efficiently identify a minimum set of essential contextual information from user requests? (ii) How can we mitigate the degradation of user experience caused by AI ``hallucinations" in code generation? (iii) How can we design a system that creates various XR interactions via natural languages without a time-consuming iteration process?
Previous studies (see \cite{de2023llmr,giunchi2024dreamcodevr, kurai2024magicitem}) have explored integrating LLM-generated code into XR environments but faced challenges with responsiveness and high error rates due to LLM's hallucination issues. Other researchers (see \cite{huang2023real,fang2024enabling,chen2024gpt, su2023voice2action}) have developed methods to generate controlling data along with scripts for real-time animation of virtual objects or robotic arms, yet these solutions did not dive deep into the fusion with the XR environments and thus cannot deliver seamless and intuitive interactions with human users.

In this paper, we introduce \sysname (\cref{fig:teaser}), the first system designed to leverage \underline{LLM}s to generate JSON data for crafting  \underline{E}xtended \underline{R}eality worlds.
Specifically, \sysname facilitates user interaction with a virtual agent through natural language, which enables an intuitive and efficient creation of virtual objects and animations within XR environments.
\sysname accepts natural language inputs initiated by hand gestures and communicates with the remote LLM server through a multi-stage strategy, selectively adding contextual information as needed to avoid overwhelming LLMs.
Unlike previous approaches that rely on script generation and runtime compilation, \sysname utilizes LLMs to directly produce JSON data, which is then used for executing various XR tasks by dedicated modules. The combination of structured JSON data and dedicated modules mitigates the complexity of code generation -- by projecting the generation problem into constrained, compact subspaces (defined with respect to the context-relevant structured JSON data), performing generation within these subspaces, and mapping and applying the results back to the original XR space.
Compared to direct script generation, focusing on JSON data enables a well-grounded approach that not only enables smooth XR interactions but also prevents scripting errors and application crashes.
To the benefit of the structured design, our proposed system can operate without fine-tuning models or coding expertise from users.

The key design aspects of \sysname are:
First, \sysname employs a novel multi-stage process to minimize distractions from irrelevant data and reduces costs associated with pay-per-use APIs. It integrates a Context Library and an LLM Wrapper to systematically process each user request in two stages. Initially, the LLM analyzes user requests and categorizes them based on contextual properties. Subsequently, the LLM Wrapper leverages these properties to retrieve essential contextual information, which is then incorporated into a new prompt sent back to the LLM.
Second, the system requests structured JSON data as output, rather than script generation, to confine the process within constrained subspaces, enhancing responsiveness and avoiding scripting errors. By providing proper JSON schemas to LLMs, \sysname obtains high-quality structured JSON data without the need for model fine-tuning.
Lastly, \sysname introduces three modules targeted for executing different XR tasks, expanding the system's capabilities while maintaining efficiency. Each module contains distinct, minimal executable units that can be combined to handle a wide variety of XR tasks.

The main contributions of our paper are summarized as follows:
\begin{itemize}
    \item We propose a novel system, \sysname, which accepts natural language inputs from users and leverages LLMs to craft interactive XR worlds through JSON data generation.
    \item We showcase the implementation details of \sysname running as a Unity application in both link-free and linked modes and deploy it on commercial off-the-shelf XR headsets. We open source our implementation for academic and/or non-commercial use\footnote{\url{https://github.com/SNeC-Lab-PSU/LLMER}}.
    \item We conduct a preliminary user study to assess the usability and robustness of \sysname, which reveals its powerful functionality and indicates opportunities for further optimization.
    \item We perform a thorough analysis of the numerical results and interview feedback from the user study. The results indicate over 80\% reduction in consumed tokens and around 60\% reduction in task completion time compared to state-of-the-art approaches.
\end{itemize}

The remainder of this paper is organized as follows: \Cref{sec:ref} summarizes existing literature in related fields. \Cref{sec:overview} showcases the overview of our proposed system. \Cref{sec:design} presents a detailed description of the modules designed for our system. \Cref{sec:implementation} provides the software and hardware implementation of our system. \Cref{sec:study} describes the setup of our preliminary user study. \Cref{sec:results} shows the evaluation results and analysis based on our user study.
\Cref{sec:discussion} discusses the current limitation and future directions for our work, and 
\Cref{sec:conclusion} concludes the paper.

\section{Related Work}\label{sec:ref}
\subsection{LLM Agents}
To the benefit of the versatility of LLMs, many researchers (see \cite{xi2023rise, guo2024large, li2024personal}) utilized LLMs as the foundation to build AI agents that adapt to diverse scenarios. Treating LLMs as universal approximate knowledge sources with reasonable randomness, some researchers (see \cite{chen2024can,gundawar2024robust,song2023llm,kambhampati2024llms}) investigated the way to leverage LLMs to facilitate the planning or reasoning tasks. Other researchers (see \cite{singh2021pre}) tried to integrate LLMs into gaming scenarios, mimicking human players to interact with virtual worlds with sufficient environmental understanding. 
Besides single agents, some researchers (see \cite{park2023generative, gong2023mindagent, abdelnabi2023llm}) considered the collaboration of multiple agents powered by LLMs, targeting believable simulations of human behavior with proposed frameworks.
However, these are not designed for XR environments that are characterized by natural interactions among human users, virtual content, and real-world environments.

\subsection{LLM in XR}
The fusion of LLM with XR environments significantly facilitates human interaction within XR environments. Researchers have raised interest in how to provide seamless integration with more convenient and powerful interactions compared with specific hand gestures or menus. Authors in \cite{fang2024enabling} simplified robotic programming by enabling LLM for prompt processing and AR for visualizing the generated waypoints. DreamCodeVR \cite{giunchi2024dreamcodevr}, powered by LLM, assists users in programming VR environments by translating spoken language into code, simplifying VR development, and enhancing accessibility for users of varying technical skills. LLMR \cite{de2023llmr} facilitates the real-time creation and modification of interactive XR experiences, leveraging novel strategies and the Unity game engine to handle complex scenarios with limited ideal training data.  PaLM-E \cite{driess2023palm} incorporated real-world continuous sensor modalities into language models. MagicItem \cite{kurai2024magicitem} integrated cluster scripts into LLMs to accelerate the generation. However, all of them are code-based designs, which usually suffer from the hallucination issue of LLMs and potentially lead to crashes of applications.
Authors in \cite{huang2023real} utilized LLM to generate and control real-time animations created by parsing structured strings that define joint movements demonstrated across various models and motions. However, they did not consider the deep fusion with the XR environments while our work provides a seamless integration for better interactions with human users.

\subsection{3D Objects and Interactive Environment}
Users in XR environments often require the creation of 3D objects for an enhanced immersive experience. Recent studies have put efforts into how to dynamically create 3D objects based on users' requests.
DreamFusion \cite{poole2022dreamfusion} generated high-fidelity coherent 3D objects and scenes for a diverse set of user-provided text prompts. Subsequently, Magic3D \cite{lin2023magic3d} was introduced to create high-quality 3D models faster than previous methods such as DreamFusion. In addition to generating objects, the creation of interactive 3D environments has been further explored. Chat-3D \cite{wang2023chat} is the dialogue system for 3D scenes, which combines LLM and 3D representations to comprehend diverse instructions for 3D scenes. 3D-LLM \cite{hong20233d} performed better on capturing 3D spatial information after introducing a 3D localization mechanism to training. However, those approaches usually pose significant generation delays and are difficult to integrate into real-time systems.

\section{System Overview}\label{sec:overview}
\begin{figure}[ht]
    \centering
    \includegraphics[width=0.48\textwidth]{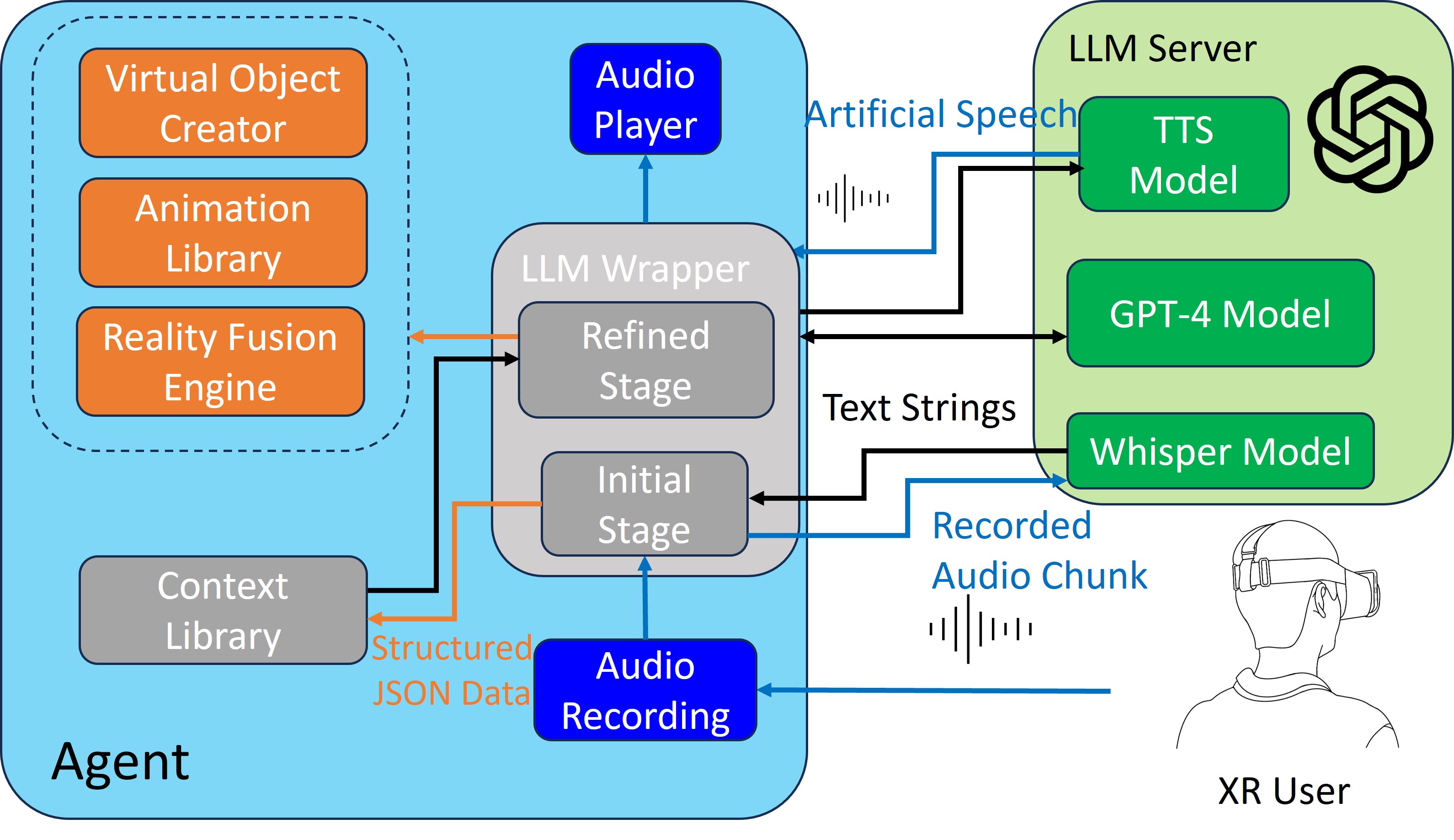}
    \caption{System architecture of \sysname.
    In the figure, blue, black, and orange lines represent the transmission of audio files, text strings, and structured JSON data, respectively. \sysname records user audio and utilizes the Whisper Model to transcribe it into text. The text-based user requests are then processed by the LLM Wrapper using a multi-stage approach, during which \sysname interacts with the GPT-4 model through text-based inputs and outputs. Additionally, the output from the LLM Wrapper can be parsed into structured JSON data or conversational text. The JSON data is used to either retrieve contextual information from the Context Library or execute various XR tasks through three designated modules, while the conversational text is converted into synthetic speech by a text-to-speech model to emulate human-like conversation.
    }
    \label{fig:arch}
\end{figure}

In this section, we provide an overview of \sysname, a system designed to craft interactive XR worlds based on users' audio commands. Utilizing state-of-the-art LLMs like GPT-4 from a cloud server, \sysname generates high-quality and structured JSON data and leverages backend modules to execute these data in accordance with user requirements. It also generates synthetic audio to emulate conversations among humans. An overview of the system architecture is presented in \cref{fig:arch}.

Unlike systems that generate scripts and compile them at runtime, \sysname employs pre-defined modules to handle various tasks and leverages JSON data to enable real-time interactions. The system comprises three primary modules to execute the generated JSON data for various XR tasks: i) Virtual Object Creator, which can generate new virtual objects using either Unity primitives or existing local prefab resources; ii) Animation Library, which produces a variety of animations in response to user requests; iii) Reality Fusion Engine, which enables the interaction between human user and virtual objects overlayed on the real-world environments. By leveraging clearly defined JSON data, \sysname avoids the error-prone script generation and accelerates the processing time compared with images or 3D meshes creation.

To minimize the distraction from the complex XR contexts and enhance the quality of the generated JSON data, we introduce two essential components: the LLM Wrapper and the Context Library. Each user request undergoes two main phases within the LLM Wrapper: \textit{initial stage} and the \textit{refined stage}. Initially, \sysname submits the raw user request, encapsulated in an \textit{initial prompt} that includes a JSON schema, to the cloud server. The server classifies the request and identifies necessary contextual categories as the \textit{initial response}. Following this, \sysname consults the Context Library to extract necessary contexts from complex XR contexts space and integrate them into \textit{refined prompts}. These refined prompts guide the generation of \textit{refined responses}, which consist of JSON data that complies with the specified JSON schema and is then utilized by aforementioned modules for executing XR tasks. For more complex inquiries, the initial stage decomposes the original request into simpler segments, producing multiple refined prompts.
Taking the example as shown in \cref{fig:teaser}, \sysname first analyzes the initial prompt and splits the user request into an object creation task and an animation generation task in the initial response. Then, it generates two different refined prompts with necessary contextual information. The refined responses from the LLMs will include structured JSON data that can be converted to desired behavior in the XR world.
Different from prior iterative approaches (e.g., the Builder-Inspector structure in LLMR \cite{de2023llmr}) focusing on inspecting and regenerating results, which often leads to indeterminate iterations with significant delays, our multi-stage process operates in a more controlled and structured manner. Moreover, this design effectively excludes unrelated contextual information and thus alleviates the distraction of LLMs.

To mimic natural human communication, \sysname includes a virtual avatar representing an AI agent in the XR environment. This avatar accepts raw audio inputs from the user and responds with synthesized speech of concise descriptions. User voice recordings are activated by hand gestures and transcribed into text using OpenAI's Whisper model \cite{radford2023robust}. Meanwhile, selected text responses from the LLM are processed through OpenAI's text-to-speech (TTS) model and delivered in artificial speech together with other interactions.
In the next section, we will present the detailed design principles for our system.

\section{System Design}\label{sec:design}
In this section, we will showcase the core design of \sysname. We begin with the description of the Context Library, focusing on its roles in generating high-quality JSON data while maintaining low cost. This is followed by the presentation on the three task modules, detailing what kind of JSON data is required and how they convert the JSON data into real-time interactions. Lastly, we introduce the LLM Wrapper, which connects all those components. 

\subsection{Context Library} \label{sec:context-lib}
Contextual understanding is crucial for seamlessly integrating LLM into XR scenes. However, complex contexts can distract LLM models from accurately extracting useful information and increase the operational costs of using pay-per-use models like the GPT-4 API. LLMR (see \cite{de2023llmr}) addresses this challenge with a Scene Analyzer module that condenses complex scene context into a succinct summary. While effective, this approach introduces additional processing latency and consumes significant token inputs. In contrast, \sysname utilizes a Context Library that supplies only the essential information needed based on the user’s request, enhancing the system's ability to interpret vague commands and interact with XR environments without consuming excessive tokens.

\begin{figure}[ht]
    \centering
    \includegraphics[width=0.48\textwidth]{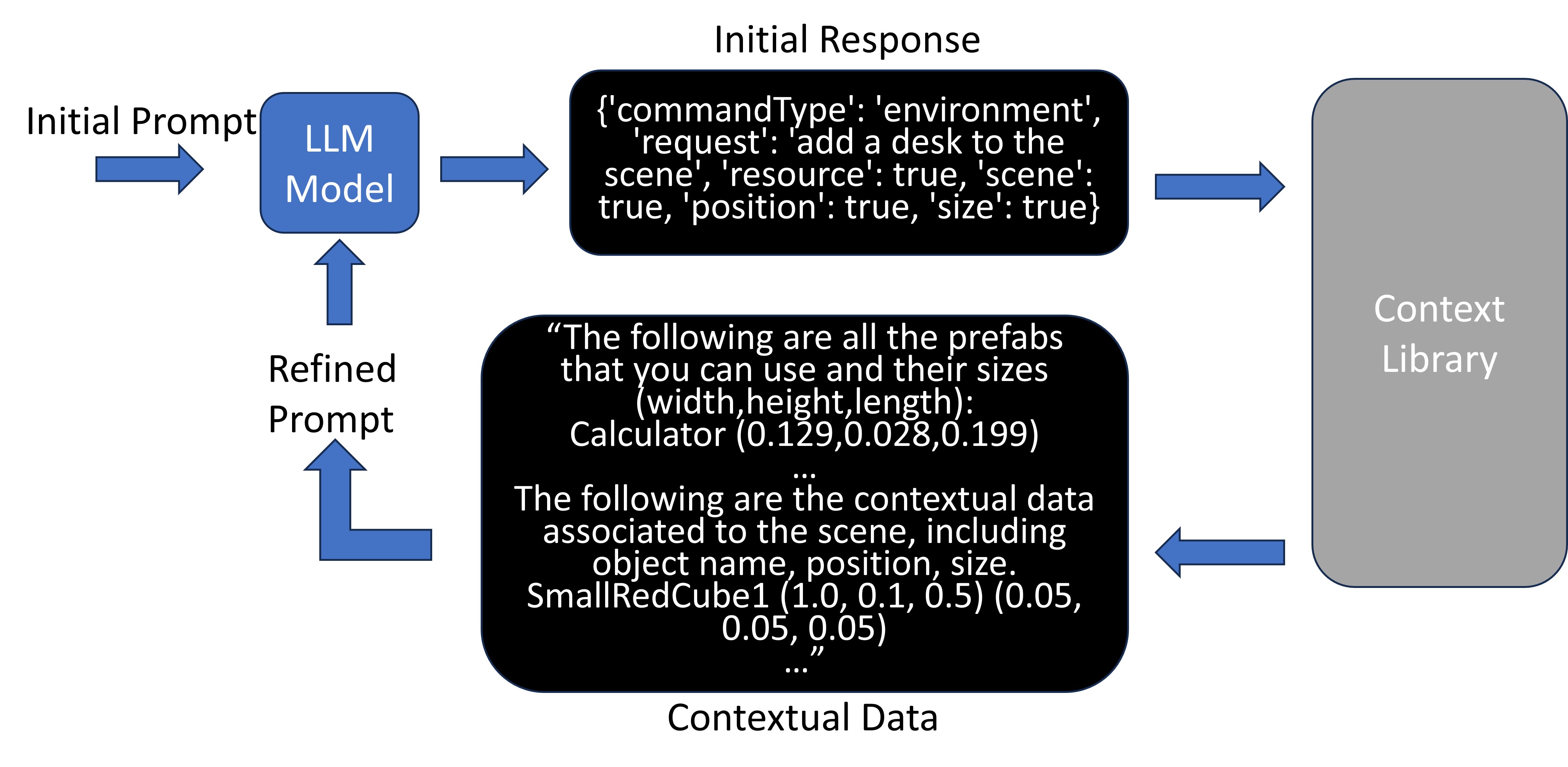}
    \caption{Example usage of the Context Library. Based on the JSON data included in the initial response, the Context Library provides essential data to be integrated into the refined prompt.}
    \label{fig:context-lib}
    \vspace{-0.1in}
\end{figure}
Within \sysname, the Context Library is accessed via the LLM Wrapper, where each user request is integrated into an initial prompt that includes a specified JSON schema. This schema directs the AI model to generate JSON data that retrieves contextual information from the Context Library. We categorize XR scene context into several crucial areas: resources for environment creation, existing scene data (distinguishing between virtual objects and real-world contexts), ongoing animations, and user-related contexts (such as tracking head and hand poses). Each category has associated properties like position, orientation, scale, and size. The LLM model dynamically selects the relevant categories and properties based on the user's request and includes them in the initial response as structured JSON data. 

The initial response's JSON data acts as an index for the Context Library to locate and provide relevant data with appropriate descriptions for the refined prompt. This design enables \sysname to request information as needed, avoiding distractions from irrelevant details and excessive token usage. An example of utilizing Context Library is shown in \cref{fig:context-lib}. 
In this example, the initial response contains required contextual categories in JSON format to place a desk on the environment, based on which the Context Library provides detailed contextual information regarding available resources and necessary environmental knowledge. This information will then be incorporated into the refined prompt for the next stage of communication with LLMs.

Historical context is also crucial, especially when the user engages in a series of closely related actions. For example, after creating a virtual object, a user may refer to it using pronouns like `it' when planning to modify its properties. To facilitate understanding of such historical contexts, \sysname maintains a queue with a fixed capacity $M$ to store the user’s previous messages. Specifically, the latest $M$ messages from the users will be recorded in text sequences, while outdated messages will be discarded to avoid overwhelming the LLM model. These recorded messages are then incorporated into the Context Library and included in both the initial and refined prompts, enabling the system to accurately interpret and respond to vague commands from the user. Throughout our experiments, $M$ is set to 10, based on the observation that users rarely refer to interactions occurring more than 10 rounds in the past.

\subsection{Virtual Object Creator} \label{sec:obj-creator}
One core functionality of \sysname is to add virtual objects to the XR scene. By default, \sysname supports creating objects from two primary sources: existing 3D models stored within the project resources and Unity primitives. It is also feasible to obtain online free models in real-time, as demonstrated in Section 4 of LLMR (see \cite{de2023llmr}). As the extension deviates from the main contribution of this paper, we left it as an optional module for future development. 

\begin{figure}[ht]
    \centering
    \includegraphics[width=0.45\textwidth]{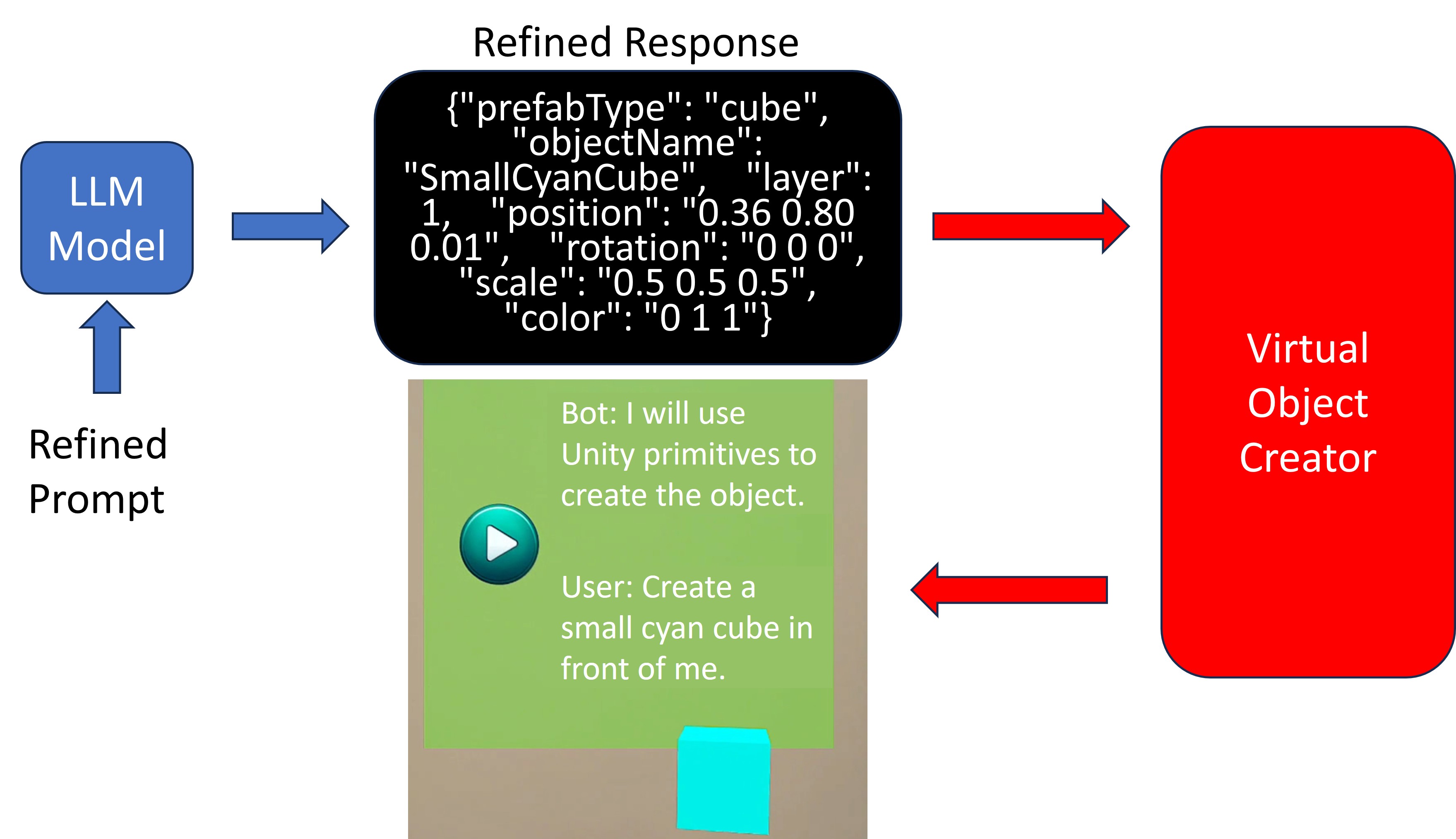}
    \caption{A request sent to the Virtual Object Creator. For requests related to creating virtual objects, the refined response is supposed to include JSON data following specific JSON schema.} 
    \label{fig:obj-creator}
\end{figure}

\begin{figure}[hbtp]
    \centering
    \subfloat[Sports car.]{\includegraphics[width=0.23\textwidth]{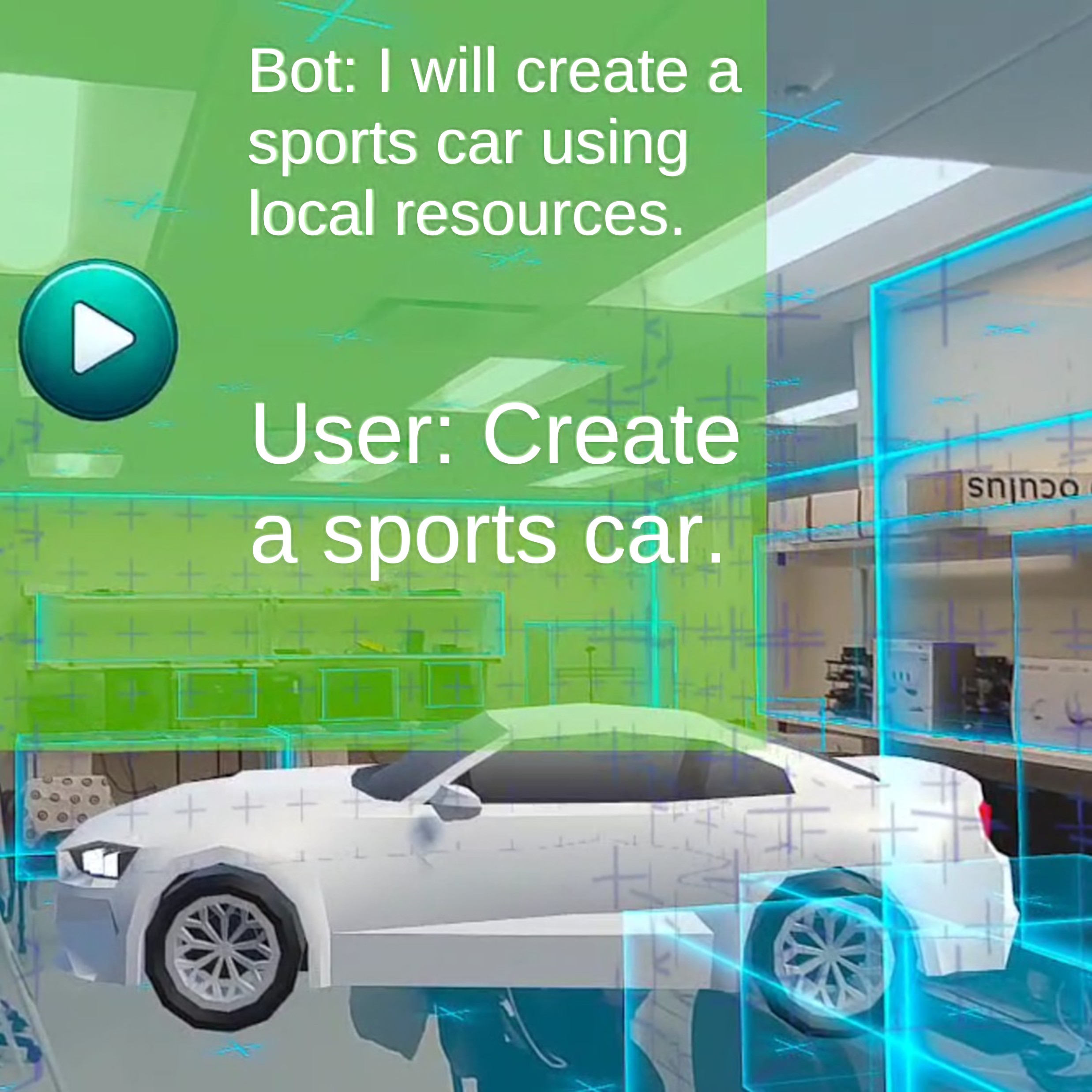}}
    \hfill
    \subfloat[Table tennis table.]{\includegraphics[width=0.23\textwidth]{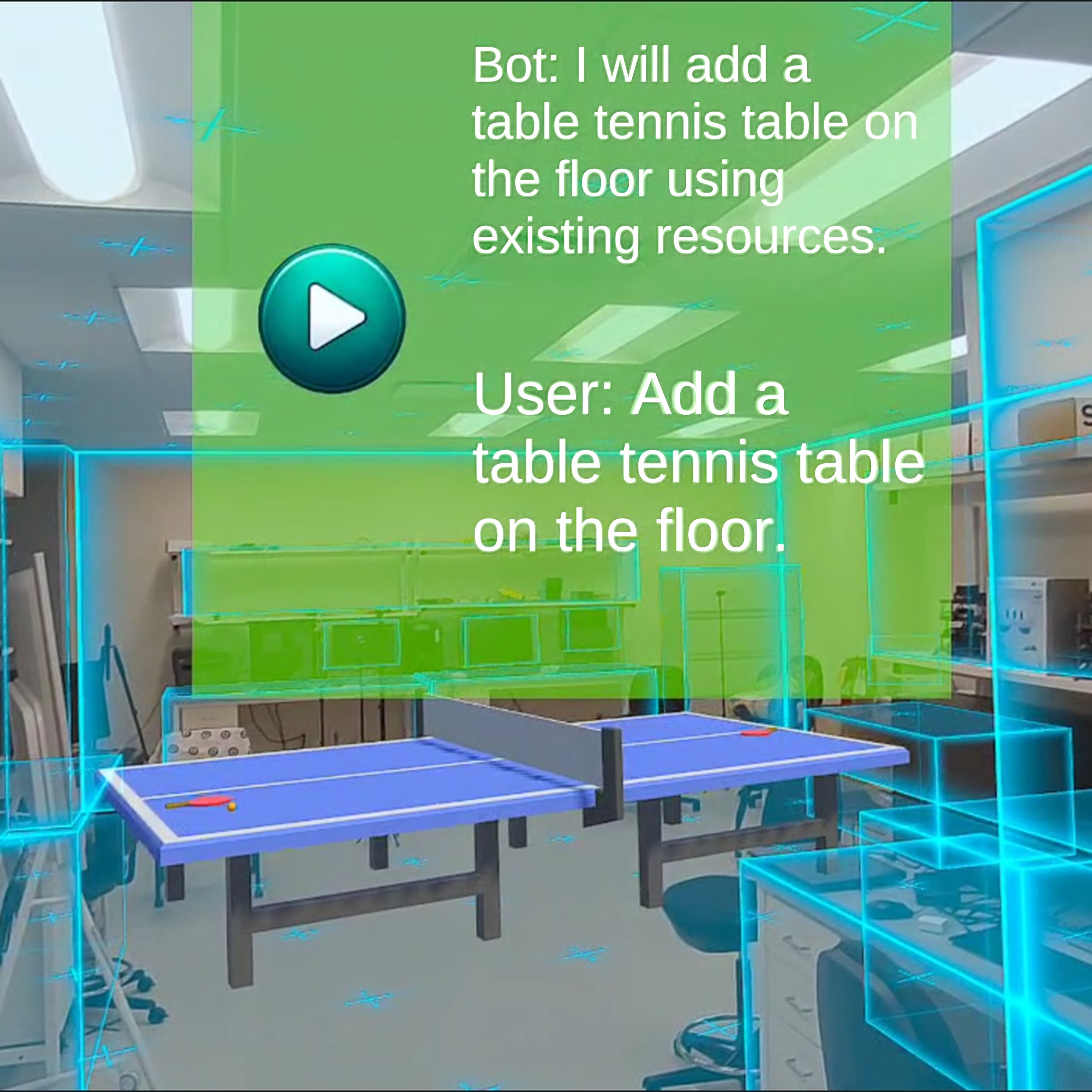}}
    \caption{Examples of creating complex objects using local resources.
    }
    \label{fig:complex-creation}
\end{figure}

As discussed in \Cref{sec:context-lib}, the refined prompt includes contextual data, such as the names of available prefab resources in the project. By analyzing this data alongside the user's request, the LLM model determines the source for object creation. Typically, it prefers using available prefabs unless otherwise specified by the user. If no appropriate local resource is available, the model generates JSON data to construct the target object using Unity primitives. The refined response for such requests typically includes the prefab name, the created object's name, and other standard properties like position, orientation, color, scale, etc., most of which are optional. The Virtual Object Creator uses this JSON data to execute backend scripts that add virtual content to the scene. An example of creating a simple cube using the Virtual Object Creator is illustrated in \cref{fig:obj-creator}. Besides creating simple objects from Unity primitives, we also provide examples of complex objects, e.g., the sports car and the table tennis table, using local resources, as shown in \cref{fig:complex-creation}. Note that the object name serves as the primary identifier for most interactions within \sysname. Therefore, we prompt the LLM to generate a descriptive name that incorporates distinct information, facilitating easier identification and interaction.

To mitigate the `hallucination' issue often associated with LLMs, especially in objects composed of multiple parts (e.g., creating a car using Unity primitives may involve multiple cubes and spheres in various sizes), we introduce some special properties. For instance, the `parent' property allows the LLM to determine dependencies between objects and generate positions in local coordinates relative to the parent. This setup also facilitates rendering checks against physical constraints, with a post-processing step to adjust positions if violations (e.g., an office supply item floating on air or placed outside the table's boundary) are detected. Notably, this check excludes empty placeholder objects used solely for coordinate referencing.

Additionally, we integrate Unity's physics engine through an optional `physics' property. Activating this property adds a rigid body to the object, if absent, and enables the use of `gravity' feature to prevent unnatural phenomena like floating objects. This property can be deactivated, particularly for objects created from Unity primitives where physical interactions are not expected. The LLM model dynamically manages these properties based on user requests to enhance the immersive experience in the XR environment.

\subsection{Animation Library} \label{sec:ani}
Static environment deviates from the immersive experience pursued in most XR applications. Thus, another fundamental functionality of \sysname is to dynamically add animations to objects within the scene. Users can request actions such as movement or rotation of specified objects. \sysname supports these interactions through its Animation Library, which maintains a series of predefined animation units that can be combined to create complex animations.

The Animation Library functions similarly to the Virtual Object Creator, transforming JSON data from refined responses into real-time animations within the XR scene. Our integration with the Unity animation system allows the following animation units:

\begin{mdframed}
\begin{itemize}
    \item Translate: Move an object to a target position.
    \item Rotate: Change an object's orientation or rotate it around an axis by a specified degree.
    \item Gaze: Let an object's orientation change with the position of another object.
    \item Orbit: Rotate an object around another.
    \item Scaling: Increase or decrease an object's size.
    \item Coloring: Transit an object's color towards the target.
    \item Attach: Designate an object as a child to another, recording the previous parent for potential detachment.
    \item Detach: Revert an object's parent to its previous state or set it to null.
    \item Catch: A complex sequence combining translation, rotation, attach, and detach, designed for the virtual avatar to mimic interactions like bringing objects somewhere.
    \item Stop: Cease a running animation.
    \item Destroy: Remove an object from the scene.
\end{itemize}
\end{mdframed}
Note that these animation units can be used individually or flexibly combined by \sysname to create a vast range of interactions based on user requests. For example, in response to a request to simulate a solar system, \sysname automatically selects and applies animation units such as Rotate, Orbit, Scaling, and Coloring to meet the specified requirements. Users are not required to have any prior knowledge of XR application development or adhere to certain rules when creating animations; they simply express their requests intuitively, and \sysname analyzes and handles the remaining workflow. Additional animation units can be easily integrated into the library to accommodate different game engines or development environments. Through extensive testing, we have found that the existing animation units adequately cover the most commonly used animations in XR environments.

\begin{figure}[hbtp]
    \centering
    \includegraphics[width=0.48\textwidth]{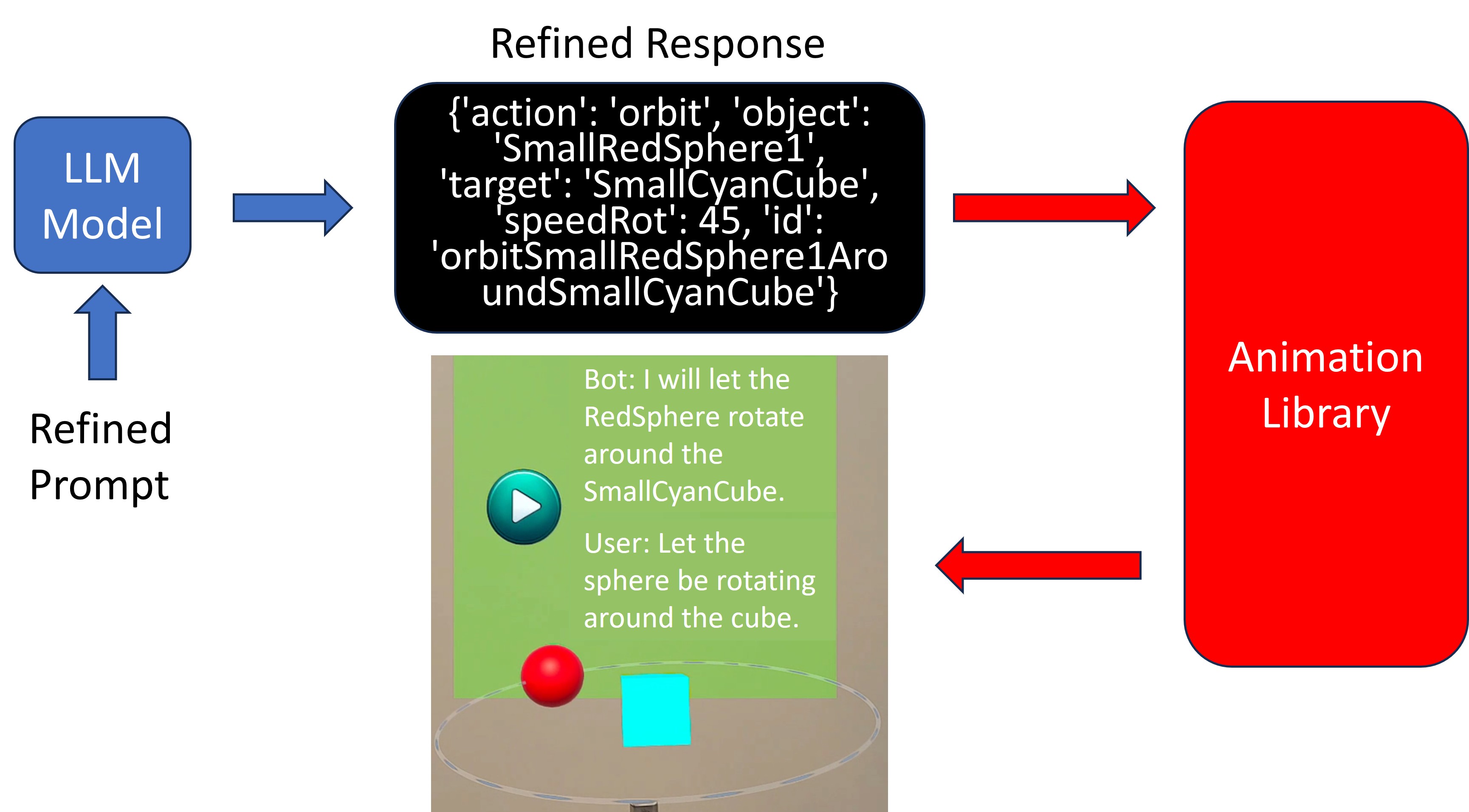}
    \caption{A request sent to the Animation Library. By parsing the action types and corresponding properties, the Animation Library generates various types of animations in a fast and reliable way.
    }
    \label{fig:ani-lib}
\end{figure}

In addition to specifying the action type, the LLM provides additional properties for more personalized animations, such as the object's name, target position, orientation, color, and attributes like speed and duration. The refined response needs to include these details, which are drawn from the context provided by the Context Library. The LLM's ability to recognize objects through vague commands, like ``the closest cube to my position," is also enabled by the Context Library, as discussed in \Cref{sec:context-lib}. \cref{fig:ani-lib} illustrates an example of calling the Animation Library during runtime.

Each animation created by the Animation Library is assigned a string ID by the LLM, facilitating a mapping mechanism within \sysname to control the execution of Unity Coroutines. These IDs, stored within the Context Library, serve as indexes to manage animations, allowing specific animations to be stopped as needed.
Moreover, the Animation Library also supports commands in both world and local coordinates, enabling the automatic conversion of local positions based on the object's transformation properties in Unity. This capability is crucial for accurately interpreting spatial commands, such as "two meters forward from its current position."
Furthermore, the Animation Library maintains an action queue and intelligently manages animation sequences based on the JSON properties and action types. For example, it can sequence a series of movements to mimic a specific trajectory or manage multiple animations running in parallel, such as rotating and orbiting motions when simulating a solar system. By leveraging Coroutines in the Unity engine, \sysname flexibly orchestrates sequential or parallel animations to fulfill user requirements.

\subsection{Reality Fusion Engine}
\begin{figure*}[ht]
    \centering
    \subfloat[Place virtual office supplies on top of the real-world table.]{\includegraphics[width=0.24\textwidth]{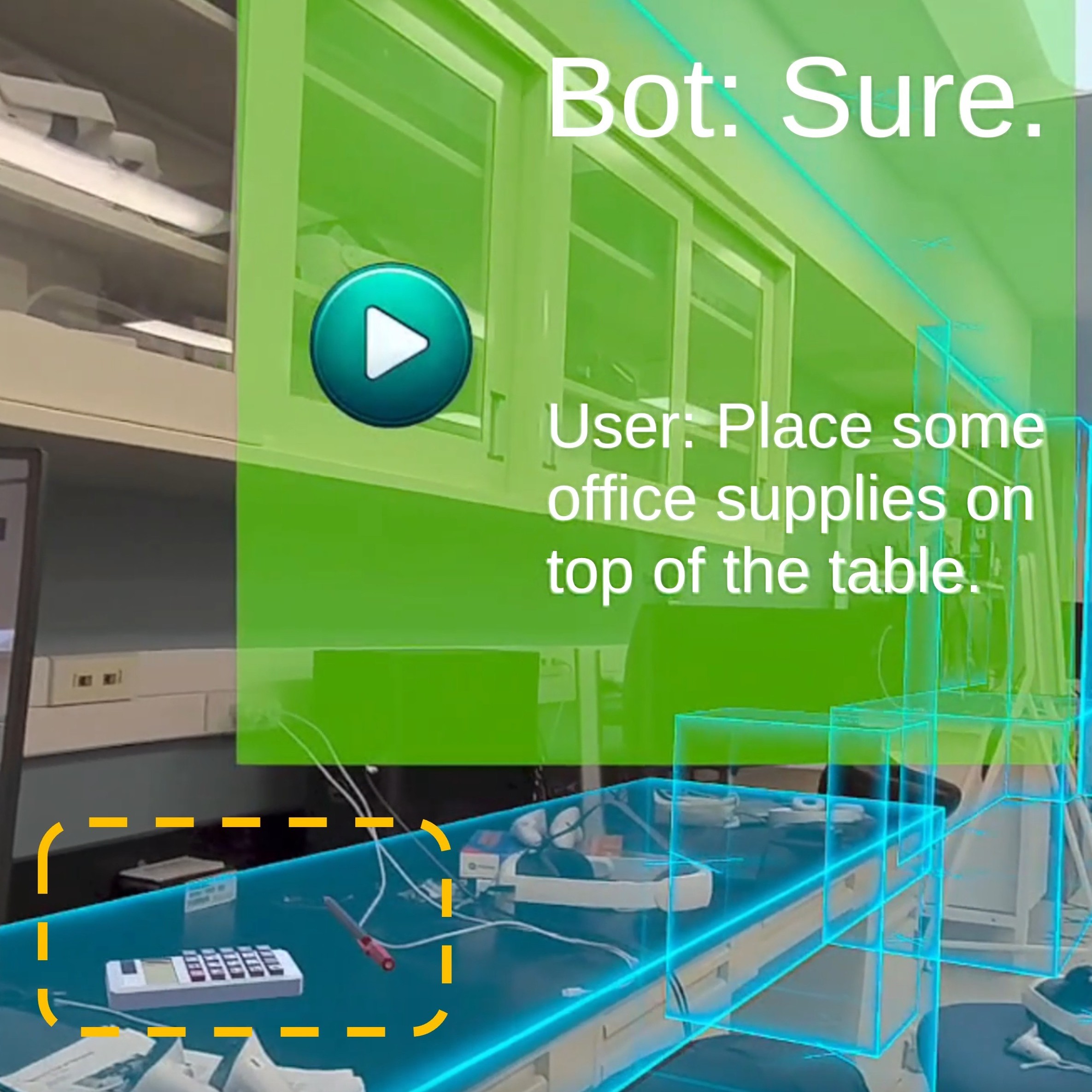}}\hspace{0.05in}
    \subfloat[Create a grabble cube.]{\includegraphics[width=0.24\textwidth]{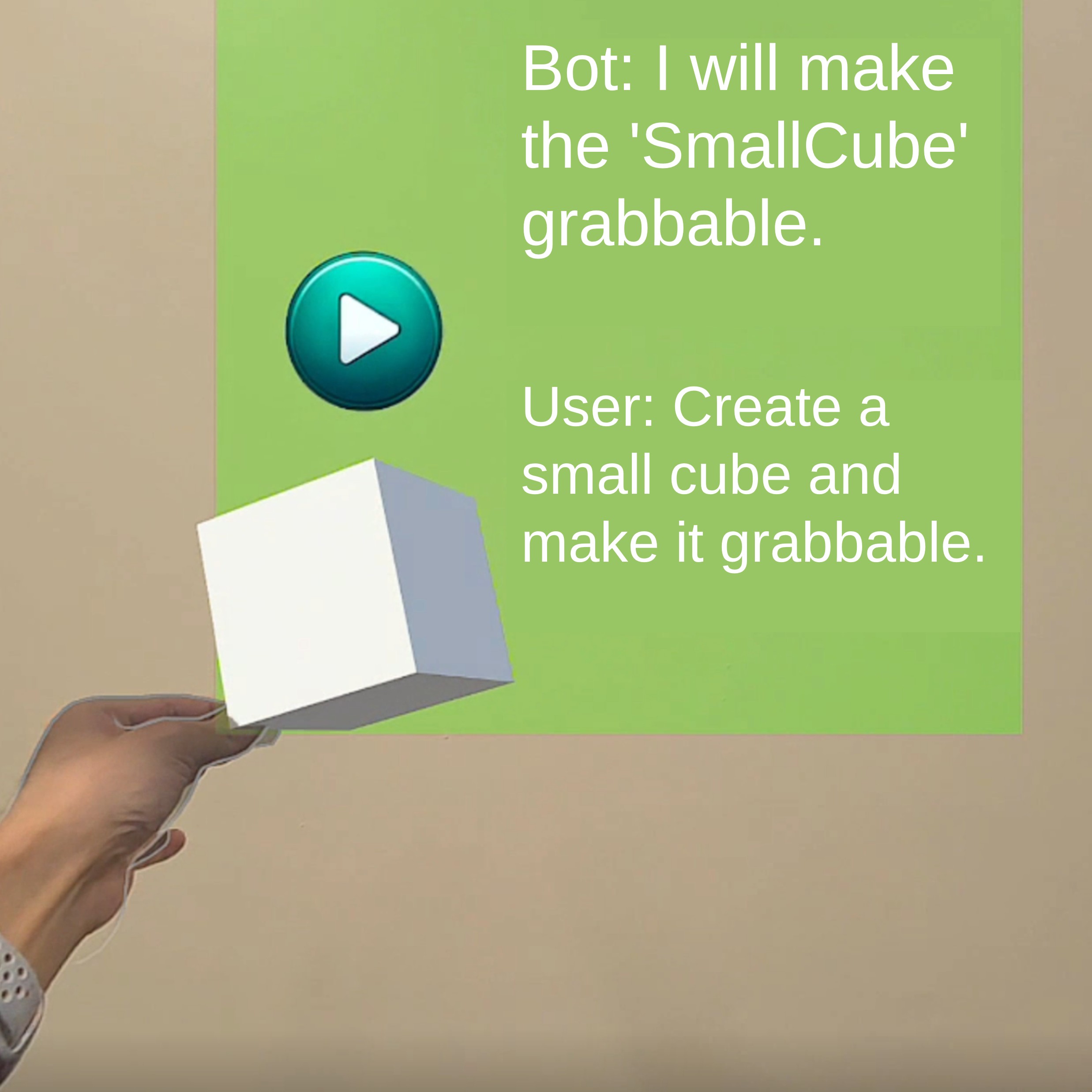}}\hspace{0.05in}
    \subfloat[Create a cube floating on top of the user's hand.]{\includegraphics[width=0.24\textwidth]{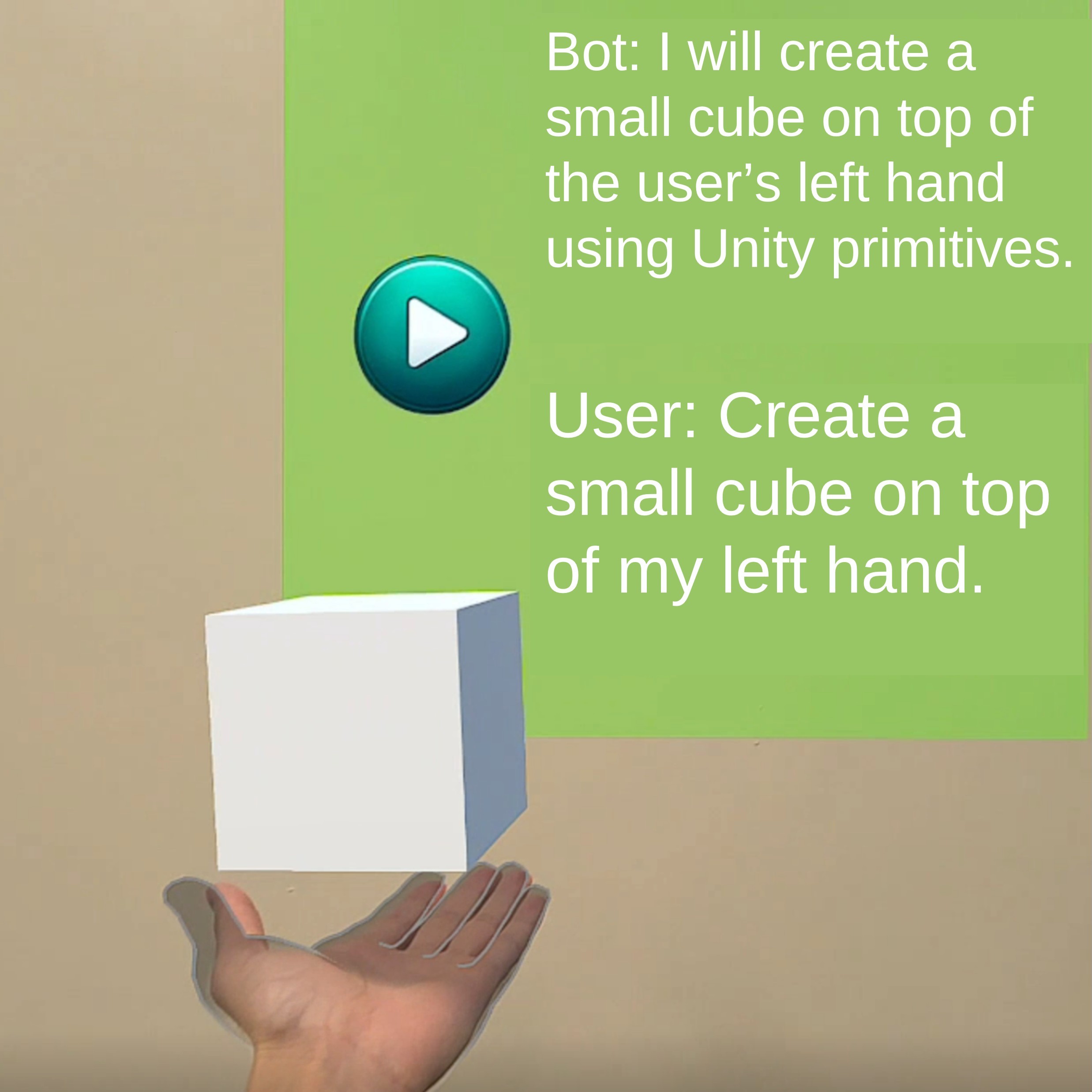}}\hspace{0.05in}
    \subfloat[Convert user drawing on whiteboard to objects.]{\includegraphics[width=0.24\textwidth]{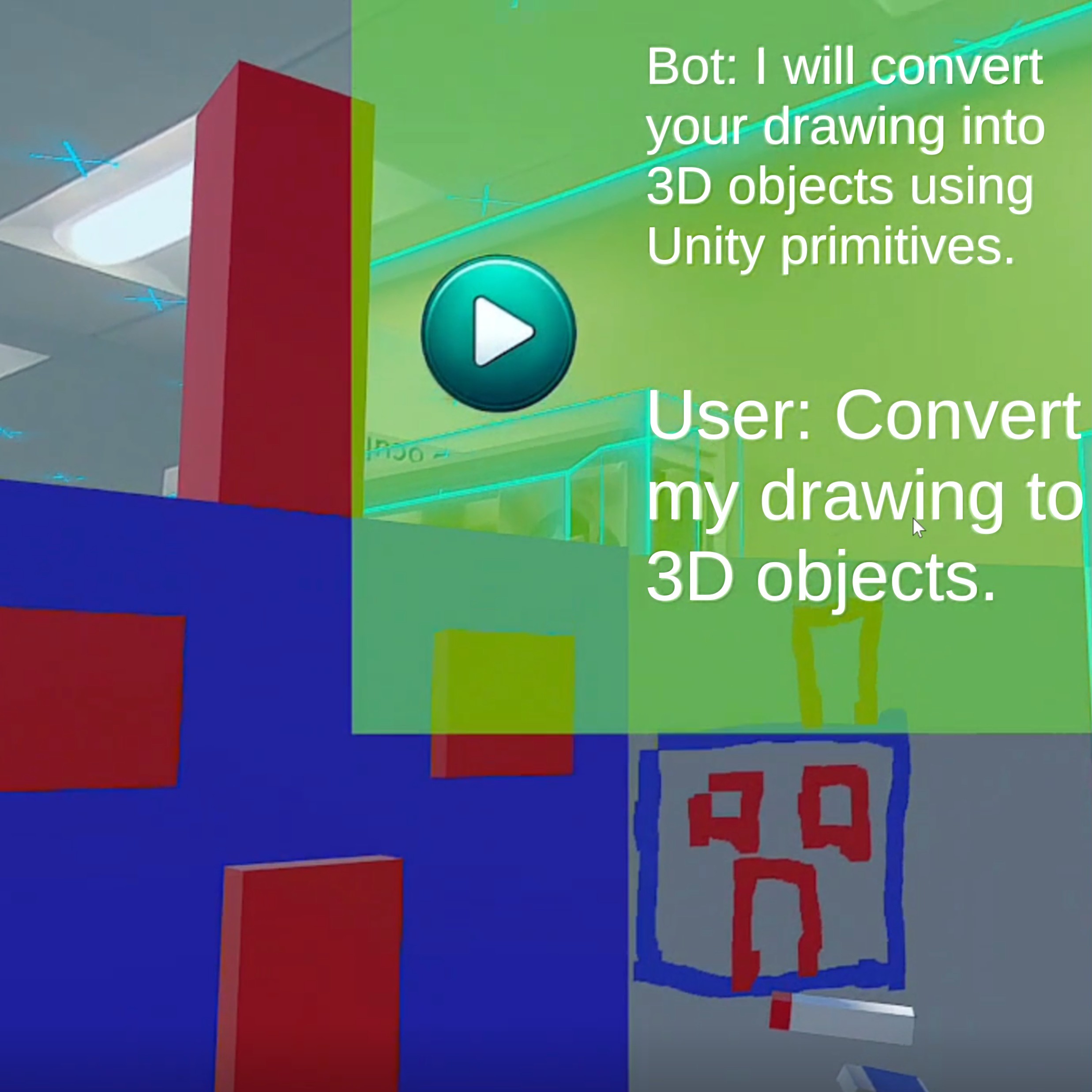}}
    \caption{Example applications powered by the Reality Fusion Engine.}
    \label{fig:ex-mrapp}
\end{figure*}

Previous work in integrating LLM with XR lacks deep fusion with real-world objects, including contextual understanding of the real-world surroundings and interaction with the user's body. \sysname explicitly addresses these limitations by introducing the Reality Fusion Engine, which seamlessly integrates real-world objects into virtual contexts and leads to more immersive interactions than traditional setups, as showcased in \cref{fig:ex-mrapp}. While the current implementation relies on Meta XR All-in-One SDK, it is designed to be adaptable to other platforms, such as Apple's RoomPlan API.

Utilizing the Scene API from the Meta XR All-in-One SDK, we assume that the user has previously scanned their environment, allowing for real-time matching with the room model. In this setup, real-world objects such as tables and storage units are represented by invisible objects equipped with precise colliders that facilitate interaction with our virtual models. These objects are assigned generic names (e.g., invisible plane, invisible volume), which prevent them from precise identification via names. However, they are tagged with descriptive labels that, along with their positions, orientations, and sizes, are cataloged in the Context Library to support immersive interactions.

Implementing XR-specific actions like picking up or throwing objects using the user's hands poses significant challenges for an LLM in generating bug-free scripts quickly. To address this, we create prefabs for the building blocks, like grabbability, which can be instantiated and attached to virtual objects to enable interaction during runtime through specific JSON data.

Furthermore, we incorporate the hand-tracking results into the contextual information, which enhances user interaction by allowing for more precise and immersive engagements with the user's body. For instance, we can place objects directly in the user's palm, with their position dynamically adjusting to the movement of the user's hand. The contextual information also provides detailed tracking for each finger's bone, enabling fine-grained operations with the fingers.

\subsection{LLM Wrapper} \label{sec:wrapper}
To seamlessly integrate various modules with the LLM, we propose the LLM Wrapper, which serves as a connector for communicating between the remote AI server and the designed modules. 

As outlined in \Cref{sec:overview}, our system utilizes three distinct models for handling various types of input and output: the Whisper model for audio transcription, the GPT-4 model for generating responses given the provided prompts, and the TTS model for text-to-speech processing. The Whisper model operates independently, processing only audio recordings. The responses from the GPT-4 model, consisting of a mix of plain text and JSON data, are segmented by punctuation to isolate plain texts, which are then forwarded to the TTS model for vocalization.

Upon receiving transcribed texts from the user's audio inputs, the LLM wrapper orchestrates the workflow through two main phases for each request: the initial and the refined stages. In each stage, it encapsulates the user's request and contextual information into predefined prompts. These prompts include a JSON schema that directs the LLM to produce appropriate JSON data. The JSON data generated by the LLM are parsed within the wrapper and utilized to trigger corresponding modules, facilitating immersive interactions in the XR environment. The parsing process can be treated as a validation process on whether the generated JSON output meets the required structure and semantics. Invalid JSON data will lead to a log of warnings and no response to the user's request, to some extent degrading the user's experience. To the benefit of our well-structured prompts and the powerful functionality of the GPT-4 model, the generation of invalid JSON data is pretty rare throughout our testing. Moreover, the latest LLMs, such as GPT-4 Omni and OpenAI o1, support structured outputs\cite{struoutput}, which ensure responses adhere to the provided JSON schema.

\section{System Implementation}\label{sec:implementation}
\sysname is built upon the Meta XR All-in-One SDK in Unity, tailored for the commercial off-the-shelf XR headsets, e.g., Meta Quest 3.
We offer two operational modes for users: one allows running the application directly from the Unity Editor by connecting the headset to a computer via cables, and the other supports deploying it to the mobile device for a link-free experience. In both configurations, users can interact with the real world and access all functionalities outlined in Sections \ref{sec:overview} and \ref{sec:design}. For user studies, we opt for the linked mode to facilitate real-time monitoring of the user status, as we will be able to provide timely help when needed.

In addition to the main application written in C$\#$ within Unity, we have developed a forward server in Python. This server acts as an intermediary to handle communications between the user and OpenAI's cloud server via the official API. It processes prompts generated by the LLM wrapper, sends them to the cloud, and relays responses back to the user. The communication between the forward server and the user is established through socket programming using the Transmission Control Protocol (TCP).

For the user study, \sysname runs on a Dell workstation equipped with an Intel Core i7-11700K CPU @ 3.60GHz, an NVIDIA GeForce RTX 3070 Graphics Card, 32 GB memory, 1 TB of disk space, and Windows 11 Enterprise. The forward server is also hosted on this machine, ensuring robust and efficient handling of data transmission via the local network.

\section{User Study}\label{sec:study}
We have recruited eight graduate students (three females, five males; mean age: 26.5 years) for our user study\footnote{All study procedures were conducted in accordance with the ethical guidelines of the Pennsylvania State University Institutional Review Board (IRB) and were approved under protocol number STUDY00024885.}, following guidelines for "debugging" tests as recommended in prior work (see \cite{bevan2003magic}). Those students are all recruited at the Pennsylvania State University through email propaganda. The participants represent diverse academic backgrounds, including computer science, electrical engineering, and materials science. Their familiarity with XR technology varies: three are novices, who are aware of XR concepts but have no practical experience; two have intermediate experience, being familiar with XR and having used XR products; and three are experts, possessing in-depth knowledge of state-of-the-art advancements in XR. This participant profile is selected to align with the target users of our proposed framework—individuals innovating in education and gaming who may have varying levels of familiarity with XR and a diverse range of backgrounds.
The primary purpose of our user study is to reveal the feasibility of leveraging LLM-generated JSON data to execute various XR tasks with simple natural language input. The study may not evaluate every facet of the system but will explore the potential and indicate the opportunities for future optimization. We will also analyze the insensitive statistics, e.g., the consumed tokens and response generation time, to demonstrate the efficiency of the proposed system.

During the study, participants can sit down or stand wearing a Meta Quest 3 HMD. They enter an XR world that maps the real-world environments. They can see a green canvas to initiate communication with \sysname, along with a virtual agent with which they can interact.
We define seven tasks for each participant to finish as follows:
\begin{enumerate}
    \item \textit{Car creation.} Create a small car and adjust the size.
    \item \textit{Office supplies placement.}  Place multiple office supplies on top of a real-world table.
    \item \textit{Cube movement.} Create a cube in some color and let the cube move towards a specific position.
    \item \textit{Solar system.} Create a simple dynamic solar system, enabling orbiting and self-rotation.
    \item \textit{Virtual agent delivery.} Let the agent catch something for the user.
    \item \textit{Get agent's attention.} Create a grabbable cube and move it to the table while keeping the agent's gaze on it. Observe how the agent's head follows the cube's movement.
    \item \textit{Recognizable drawing.} Create a whiteboard, perform a 2D drawing on it, and let the agent recognize the drawing. The user may change the color of the tip during drawing or clear the drawing by both voice and eraser. 
\end{enumerate}
Those tasks cover a range of topics with various difficulty levels. Tasks 1 and Task 2 are both 3D creations, where Task 1 creates complex objects, and Task 2 involves understanding the real world. Task 3 and Task 4 are 3D animations in different categories. Task 5 and Task 6 showcase interaction with the virtual agent. Task 7 is more advanced and involves the vision-based recognition of various drawings. Note that compared with tasks defined in \cite{giunchi2024dreamcodevr}, our tasks share some commonality but involve more complex components, aiming to evaluate the interactivity and the capability of LLM to understand real-world environments.

At the beginning of the user study, participants are provided with a concise system overview. This is followed by a 10-minute training session, allowing them to communicate with the virtual agent to execute simple tasks such as creating a cube, coloring it, and moving it. Subsequently, participants are requested to complete each task in order. We provide a combination of
methods to perform the evaluation by collecting non-sensitive information from the users:
\begin{itemize}
    \item Cost analysis. We collect the consumed tokens and response generation time for each user request as quantitative metrics to compare with baseline methods. The consumed tokens serve as a direct indicator of the cost when calling pay-per-use APIs, while the response generation time is defined as the consumed time from the time the request is sent to the remote AI server to the time of receiving a complete response. 
    \item Task-level usability metrics. We use task completion time, fulfillment, and number of help requests as metrics to evaluate the usability of our system. In particular, efficiency and effectiveness are captured through task completion time. User satisfaction is measured by fulfillment. Ease of use and intuitiveness are evaluated by the number of help requests.
    \item Questionnaire. The questionnaire gathered all users' insights: usability, immersion, interactivity, future use, and overall experience. The questionnaire design is based on existing surveys \cite{lewis1995ibm,lund2001measuring}. After the study, each participant is required to fill out this seven-question questionnaire on a seven-point Likert scale about their experience using \sysname. The scale ranged from 1 (strongly disagree) to 7 (strongly agree). The statements included in the questionnaire are shown as follows:
\begin{enumerate}[label={(Q\arabic*)}]
    \item I felt immersed in using this system.
    \item I like the interaction possibilities enabled by this system.
    \item The system was responsive and reacted to my commands in an acceptable time.
    \item It is simple to operate the system.
    \item What is produced by this system meets my expectations.
    \item I would like to use this system in the future.
    \item I am pleased with the overall experience.
\end{enumerate}
    \item Semi-structured interview. We design some open-ended questions to encourage users to express their experience after they finish all the tasks. They are also flexible to express their own opinions which are not covered in the designed semi-structured interview. The questions in the semi-structured interview are listed as follows:
\begin{itemize}
    \item How easy was it to use \sysname? Could you intuitively understand the controls?
    \item Have you ever used any tools to create or manipulate 3D objects in XR or other platforms? How was this approach similar to or different from others?  
    \item What were some surprising parts of this framework? 
    \item What features or functionalities do you find most useful in this tool? 
    \item What met your expectations? What did you think could be done but didn't? 
    \item What factors would influence your decision to adopt \sysname for further use in your work? 
\end{itemize}
    
\end{itemize}

\section{Results and Analysis}\label{sec:results}
In this section, we perform a comprehensive analysis based on non-sensitive data collected during the user study. Specifically, we analyze the consumed tokens and various task completion metrics as quantitative indicators for comparison with other approaches. Additionally, we examine questionnaire scores and interview responses to assess the usability of the proposed system and explore opportunities for further optimization.

\subsection{Cost Analysis}
Recall that we collect the token consumption and response generation time for each user request in \sysname. Note that users may submit multiple requests per task until satisfied; however, for a fair comparison with existing works, we present single-request metrics here, with task-level metrics discussed in subsequent subsections.
On the other hand, \sysname does not compile code during runtime, resulting in a zero error rate as defined in \cite{de2023llmr} and \cite{giunchi2024dreamcodevr}. Thus, no error rate measurements are necessary for \sysname, as it inherently eliminates runtime compilation errors typical in similar technologies.

On average, \sysname consumes about 3,200 input tokens and 80 output tokens per request, totaling less than 3,300 tokens. This results in over 80\% reduction compared to around 30,000 tokens required per request by running similar requests on the open-sourced version of LLMR (\cite{de2023llmr}). The reduced token consumption in \sysname can be attributed to the strategic design of the Context Library and the use of JSON data generation rather than script creation.

Additionally, the average response generation time for tasks in \sysname is only 10.35 seconds, substantially quicker than the 20 to 90 seconds reported for various baseline models in LLMR (\cite{de2023llmr}). This improved response time is likely due to the lower token count and the compact subspace of generating JSON data over coding scripts. Although the requests compared may not be identical, they are similar enough at a high level to validate the significant performance enhancements achieved by \sysname.

\begin{table}[ht]
\centering
\begin{tabular}{@{}lcccc@{}} 
\toprule
Item & Average (sec) & Standard Deviation (sec) \\
\midrule
Task 1        & 89.0       & 64.80      \\
Task 2       & 75.12      & 40.48  \\
Task 3      & 69.37      & 41.14   \\
Task 4      & 157.2      & 53.80  \\
Task 5      & 74.0      & 63.95  \\
Task 6      & 131.75      & 83.87 \\
Task 7      & 160.37      & 63.33 \\
\bottomrule
\end{tabular}
\caption{Task completion time statistics.}
\label{tab:model_performance}
\end{table}
\begin{figure}[hbtp]
    \centering   \includegraphics[width=0.43\textwidth]{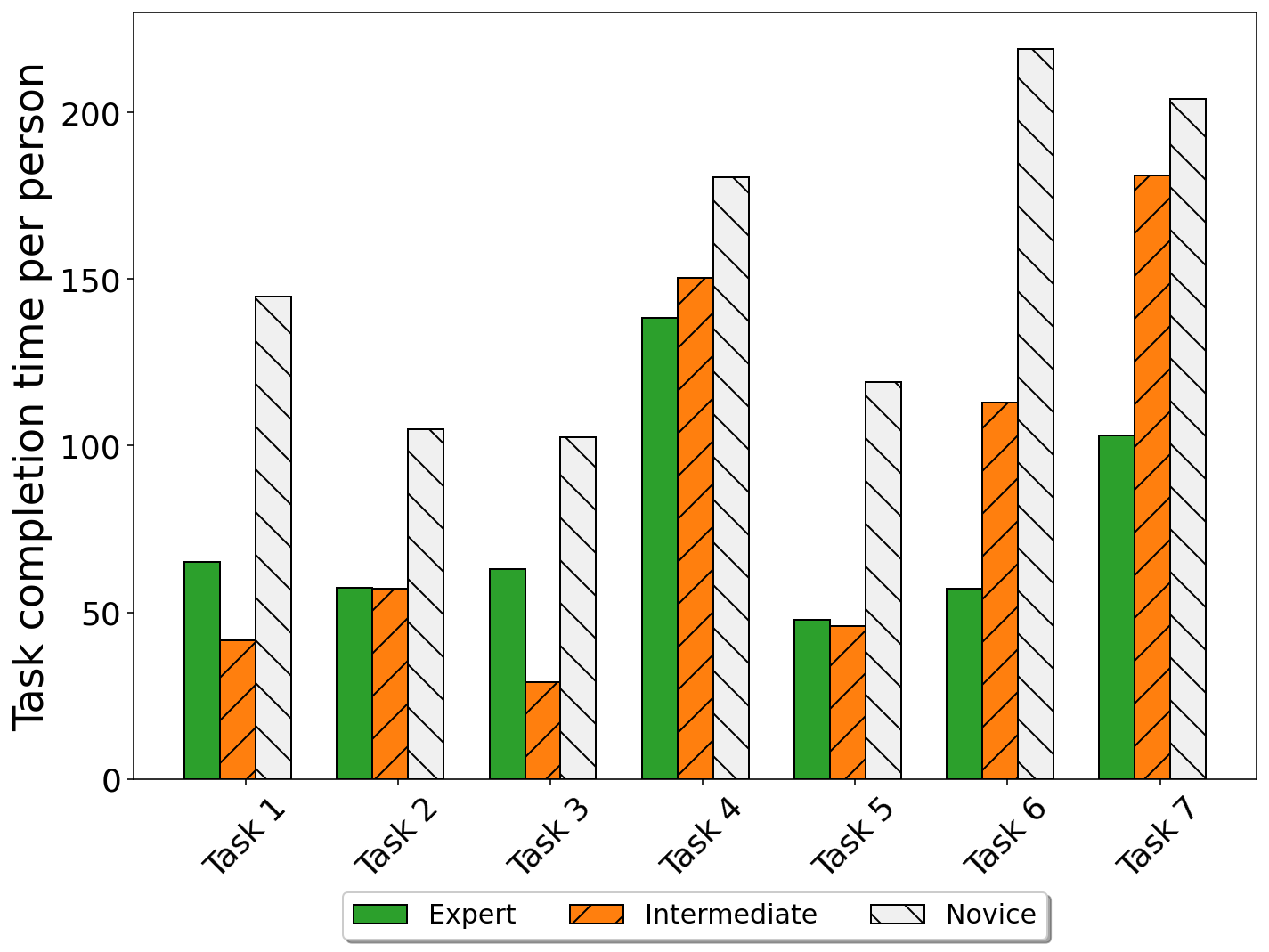}
    \caption{Task completion time regarding XR familiarity.}
    \label{fig:task_completion_per_person}
\end{figure}
\subsection{Subjective Usability Metrics}
\textit{Task Completion Time}. 
The task completion time is measured from the moment the participant begins the task to when they indicate its completion. The average value and standard deviation of all participants for each task are shown in \Cref{tab:model_performance}. Intuitively, it will be much longer than the response generation time of LLMs as it also includes the user's actions and thinking time. Besides, the deviation is mainly caused by the variations in user behavior. For instance, when the task is complicated, some users tend to use one sentence to describe their request, while some users like to split the request into several small requests, thus resulting in more rounds for our system to finish the task. Recall that we design these distinct tasks to encompass a range of XR activities and demonstrate the efficiency and versatility of \sysname. While the tasks are not identical to those in previous works, some share similarities, allowing for quantitative comparisons. In particular, our Task 3 and Task 4 closely resemble Task 1 and Task 3 from DreamCodeVR \cite{giunchi2024dreamcodevr}, respectively.
However, the average completion time of our system (69.37s/157.2s) is much shorter than corresponding tasks in DreamCodeVR (167.7s/271.4s), implying a time reduction of about 60\% in both basks.
The significant improvement in the task completion time is attributed to the reduced response generation time, thanks to the more compact subspace of JSON data generation compared to code generation. Additionally, the robustness of the proposed system eliminates the need for users to clarify similar requests repeatedly, resulting in substantial time savings.
We also analyze the average task completion time by categorizing users based on their familiarity with XR technology, as illustrated in \cref{fig:task_completion_per_person}. As expected, individuals with prior XR experience completed tasks more quickly than those with no prior exposure to XR devices. This is primarily because experienced users spent less time initiating interactions with virtual agents through hand gestures.
Furthermore, the average task completion times for intermediate and expert users are comparable across most tasks. This suggests that once users acquire basic knowledge of XR products, they encounter no significant additional challenges. However, differences become apparent in handling more complex tasks (e.g., Task 6 and 7), where experts naturally outperform intermediate users due to their richer experience and expertise in navigating such challenges.

\begin{figure}[hbtp]
    \centering
    \includegraphics[width=0.48\textwidth]{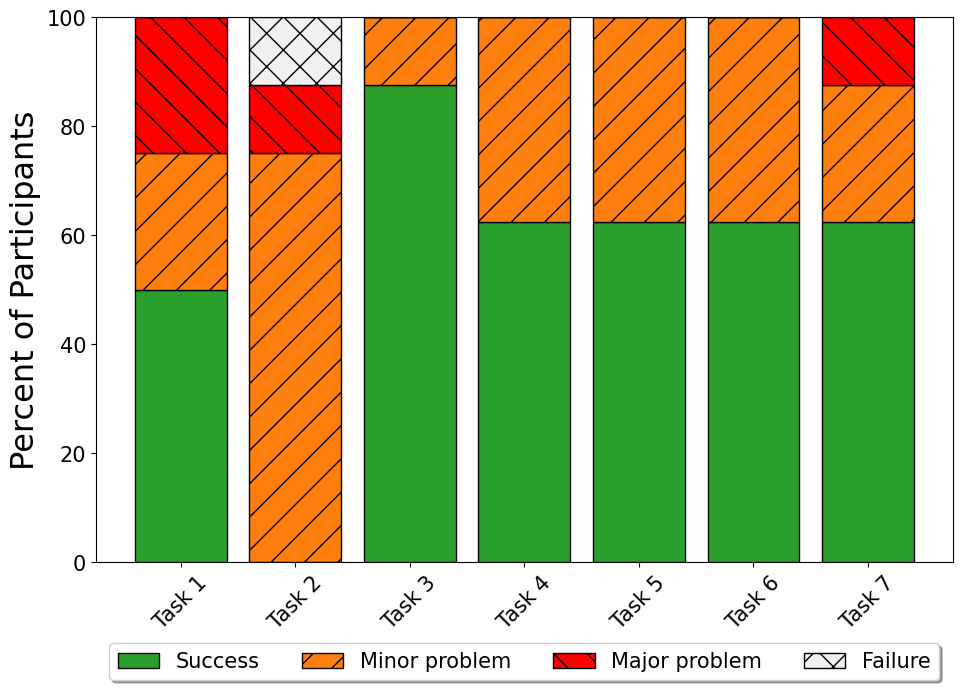}
    \caption{Fulfillment levels.}
    \label{fig:fulfill}
\end{figure}

\textit{Fulfillment}. After completing each task, users will rate the outcome based on four levels of fulfillment: success, minor problem, major problem, and failure.
Specifically, we define success as the generated results perfectly aligning with the task goals. Minor problems refer to small discrepancies, such as objects being slightly misaligned, shifted to a slightly different position, or rendered with slightly counterintuitive sizes. Major problems occur when the generated objects significantly deviate from the expected criteria, such as when the shapes of objects or animation behaviors differ substantially from what was intended.
Finally, a failure occurs when the system does not respond or execute entirely incorrect commands.
The distribution of fulfillment levels for each task is shown in \cref{fig:fulfill}. Apart from Task 2, most participants viewed the other tasks as successful. Despite being a straightforward 3D creation, Task 2 requires precise understanding and detection of the real-world environment. For instance, office supplies are sometimes inaccurately placed—either on the floor, in the air, or not correctly on the table surface, occasionally overlapping with it.

\begin{figure}[ht]
    \centering
    \includegraphics[width=0.48\textwidth]{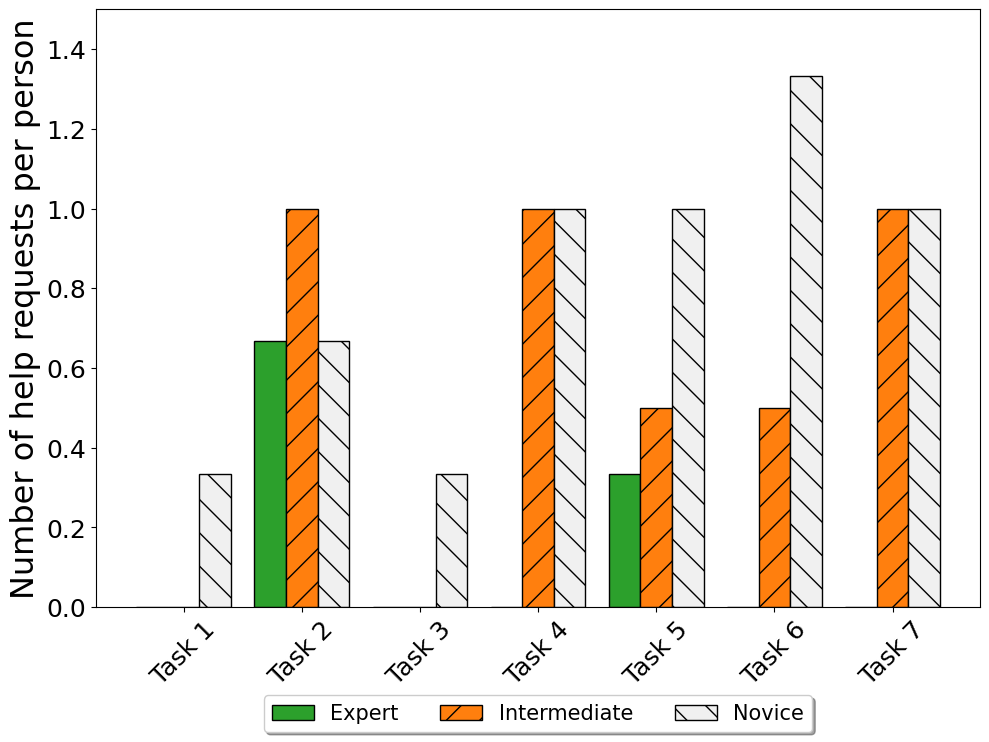}
    \caption{Number of help requests regarding XR familiarity.}
    \label{fig:help_request}
\end{figure}

\textit{Number of help requests}.
We also record the number of help requests by users when executing each task, as shown in \cref{fig:help_request}. We compare the difference in the number of requests among experts, intermediates, and novices and select the number of help requests per person as the measurement metric.
Our observations indicate that experts only requested help in Task 2 and Task 5, whereas novices required assistance for all tasks. However, the average number of help requests per person remains below $1.5$ across all tasks and user familiarity levels with XR.

\subsection{Questionnaire}
\begin{figure}[ht]
    \centering
    \includegraphics[width=0.48\textwidth]{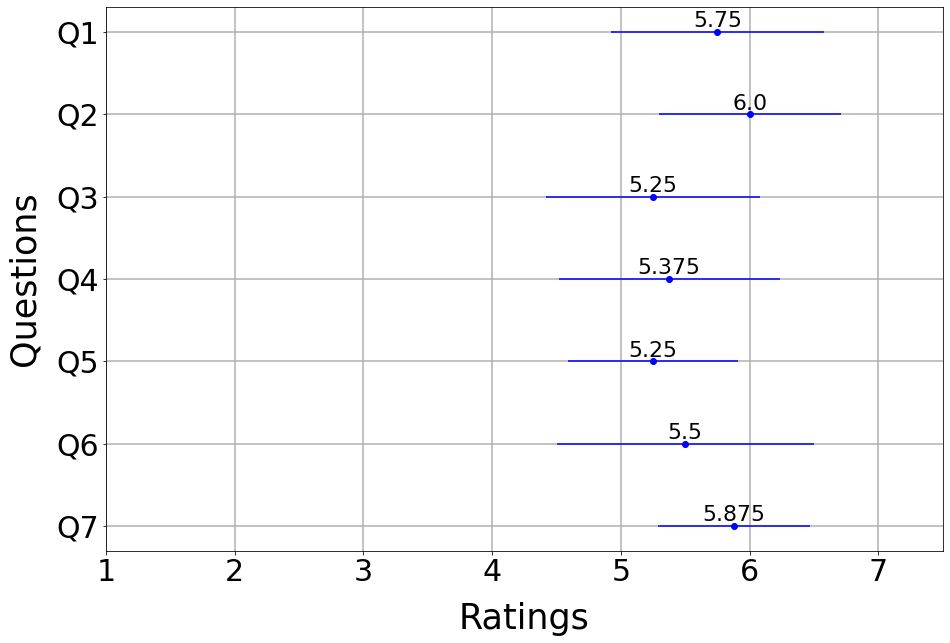}
    \caption{Results of Questionnaire. Responses were measured on a 7-point Likert scale: 1 = Strongly Disagree, 7 = Strongly Agree.} 
    \label{fig:questionnaire}
\end{figure}

As depicted in \cref{fig:questionnaire}, we can observe that all participants provided positive feedback in the questionnaire regarding usability, immersion, interactivity, future use, and overall experience (specifically, the average score of each question $\geq$ 5.25). Users are particularly impressed by the system's interactivity and immersion. They are very satisfied with the overall XR experience. In terms of usability aspects like responsiveness, ease of operation, and accuracy, the feedback is moderately positive, indicating that there is still potential for improvement to enhance user satisfaction.

\subsection{Semi-Structured Interview}
\subsubsection{User Satisfaction and Surprise} 
 Participants value the intuitive controls and the realistic interactions, which enhance their sense of presence within the XR environment. Many report a deep sense of immersion. Novice XR users find the experience particularly surprising and new, especially enjoying the engaging Tasks 5 to 7. One user remarks,``\textit{I love this system's imagination and possibilities. In Task 7, although my drawing wasn't great, the system impressively recognized and accurately transformed it into an object.}''. Another user highlights his surprise to interact with the virtual agent (virtual robot) set in XR and comments,``\textit{ It's fascinating to interact with the robot so immersively, especially being able to control where the robot focuses its attention.}''.

\subsubsection{User Critiques and Areas for Improvement}
Many users have criticized the system for lacking the capacity to learn about context and sometimes being memoryless. One user points out ``\textit{This system can't remember the previous request. For example, if I asked this system to create a sphere. Then, if we want to manipulate this sphere again, if I just use `it' to represent the sphere created before, the system could fail to understand that `it' represents the sphere}''. Another limitation of our system is that 3D objects created by our system can be too simple and abstract. It is a consensus from \textit{three} of the participants who all think the car created in Task 1 is abstract. This problem is caused by the limitation of assembling simple 3D objects such as cubes and spheres to create items requested by users. Additionally, the gesture to control the voice input can be ineffective for some users. One user comments ``\textit{My gestures are not easily captured by the system}''. It is time-consuming for those users whose gestures are not easily detected by the system to finish each task. Furthermore, the current system also can't fully understand the real-world environment. For example, in Task 2, many users complain it's hard to find the table accurately in the real world. Even if our system has its current limitations, our participants still have a positive attitude toward the future use of our system. One user comments ``\textit{If the system can achieve better performance on understanding context, the 3D creation and animation in XR can be applied to factory assembly.}''.

\subsection{User Study Insights}
Overall, the subjective usability metrics provide further positive feedback, with most participants rating the system highly across dimensions such as immersion, interactivity, and overall satisfaction. The average scores across all questions are consistently above 5.25 on a 7-point Likert scale. Besides, the user study yields a high task fulfillment ratio and a low number of help requests. All of those indicate the effectiveness of the proposed system, even for users without much programming knowledge or XR experience. Moreover, the statistics analysis reveals an over 80\% reduction of consumed tokens for each user request and around 60\% reduction in task completion time compared with other state-of-the-art works.

The semi-structured interviews with users also highlight satisfaction with the system's intuitive controls and immersive experience, particularly among novice users. However, we also identify some opportunities for further optimization, including the system's inability to retain contextual memory across requests and the undesirable appearance of generated 3D objects. Additionally, gesture-based inputs are noted as occasionally ineffective, impacting task completion for certain users. Despite these challenges, participants express optimism about the future potential of \sysname, particularly in real-world applications such as factory assembly, provided further improvements in contextual understanding and object creation are made.

\section{Limitations and Future Work}\label{sec:discussion}
In this section, we discuss several unresolved aspects of our current work and outline potential directions for future research.

\textbf{Impact of different LLMs.} Previous studies on code generation (see \cite{de2023llmr,giunchi2024dreamcodevr}) have relied on the capabilities of powerful LLMs to produce high-quality outputs, aiming to minimize error rates. In contrast, our system is robust against variations in LLM responses because of the compact constrained generation space. 
Nevertheless, the primary source of voice-to-action latency lies in the response generation time, which is largely determined by the underlying models. Using a lightweight LLM can significantly reduce latency, though this comes at the expense of response quality.
Given the continuous development of new LLMs in both academia and industry, future work could involve a comprehensive quantitative comparison of different models, assessing their performance in terms of response speed, accuracy, and contextual understanding within XR environments. Such comparisons would help identify the most suitable models for specific XR tasks and further optimize system efficiency.

\textbf{Creation of complex 3D objects.} There are two ways for \sysname to create virtual objects: existing 3D models stored within the project resources and Unity primitives. \sysname will first try to create objects by searching for local resources; it will use Unity primitives only if it cannot find the requested objects or the user explicitly indicates using Unity primitives. However, adding 3D models to the project resources can be cumbersome, while the combination of Unity primitives cannot yield high-fidelity 3D objects. Prior works (e.g., LLMR \cite{de2023llmr}) have proposed a way to download 3D models from online libraries in real-time, which can be potentially integrated into our framework to extend the flexibility of the virtual object creator module.

\textbf{Combining JSON generation with code generation.} Our approach projects the complex code generation problem into a more compact subspace, thereby enhancing system efficiency and minimizing scripting errors. While this method offers significant advantages, it also introduces certain limitations. The constrained subspace can restrict the system's flexibility in handling more generalized tasks, which code-based approaches are typically better suited for. Code generation methods, though slower and more error-prone, are often capable of addressing a wider range of use cases. A promising future direction could involve integrating both JSON generation and code generation strategies, dynamically selecting the most appropriate approach based on the user’s request. By combining these two methods, the system could retain the responsiveness and reliability of JSON generation for specific tasks while leveraging the broader functionality of code generation if needed.

\textbf{Scalability and adaptability.} Another area worth exploring is the scalability of our system across different hardware platforms and XR devices. As XR technologies evolve and new platforms like Apple's RoomPlan API emerge, the adaptability of \sysname to different ecosystems will become increasingly important. Future work could involve optimizing the system for cross-platform compatibility, ensuring it performs efficiently on a variety of XR setups while maintaining a high level of interactivity and immersion.

\section{Conclusion}\label{sec:conclusion}
In this paper, we introduced \sysname, a novel framework designed to craft interactive XR worlds using JSON data generated by LLMs. \sysname listens to the user's audio input, leveraging LLM to convert it into JSON data that is subsequently used to create virtual objects and animations within the XR scene. To minimize distractions from complex XR contexts and reduce the cost of calling pay-per-use APIs, \sysname incorporates a Context Library and an LLM Wrapper, which provide only the essential contextual information based on the user's request. Additionally, the system utilizes pre-defined scripts across various modules to support diverse interactions within the XR environment. We conducted an IRB-approved user study with eight participants to gather feedback on their experience using our framework. The study's results highlight the efficiency of \sysname, illuminating a new direction for creating truly interactive XR environments through LLM-powered JSON data generation, and uncover a series of opportunities for further optimization and investigation.



\acknowledgments{
The authors wish to thank all the participants in user study. This work has been supported in part by NSF under the grants CNS-2152610, CNS-2152658, M3X-2420351 and M3X-2420352, and DARPA grant HR0011-2420366.}

\bibliographystyle{abbrv-doi}

\bibliography{refs}
\end{document}